\documentclass[preprintnumbers,amsmath,amssymb]{revtex4}
\usepackage{amsmath}    
\usepackage{graphicx}   
\usepackage{color}

\pdfoutput=1
\begin{document}

\title{Using abstract elastic membranes to learn about quantum measurements}

\author{Massimiliano Sassoli de Bianchi}
\affiliation{Center Leo Apostel for Interdisciplinary Studies, Brussels Free University, Krijgskundestraat 33, 1160 Brussels (Belgium)}
\email{msassoli@vub.ac.be}


\begin{abstract} 
\noindent We provide an accessible popular science version of a research article published by Diederik Aerts and the author [Ann. Phys. 351 (2014) 975--1025], where an extended version of the quantum formalism was proposed as a possible solution to the measurement problem.
\end{abstract} 

\maketitle

\begin{figure}[htbp]
\begin{center}
\includegraphics[width=4cm]{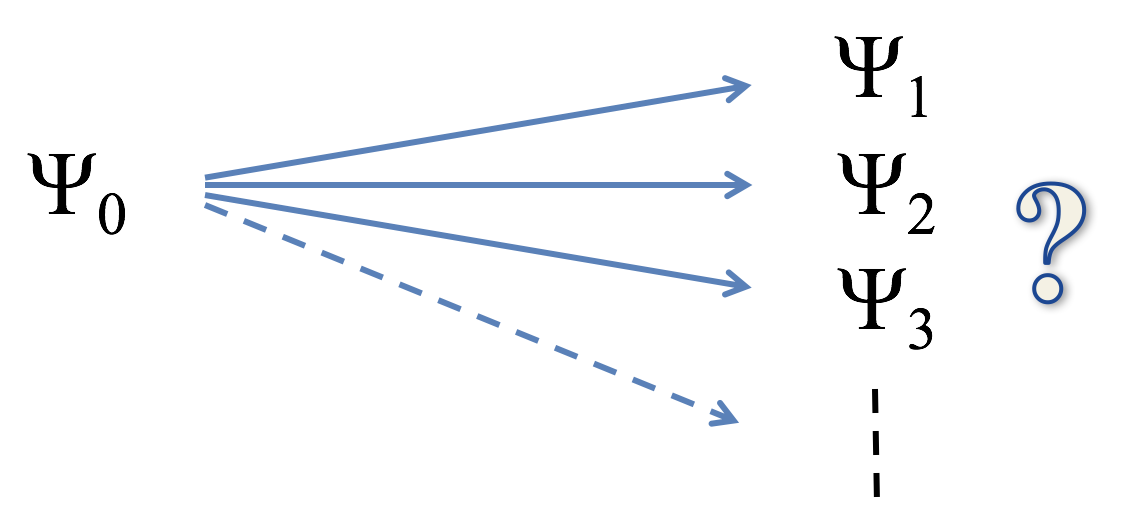}
\caption{Only one among the available outcome-states is actualized at each run of a quantum measurement.} 
\label{Figure1}
\end{center}
\end{figure}

In quantum mechanics, the \emph{state} of a physical system can evolve either deterministically (in a predictable way), when it does not interact with its environment, or nondeterministically (in an unpredictable way), when subjected to a so-called \emph{measurement process}. In the latter case, the initial (pre-mesurement) state of the physical system, let us call it $\psi_0$, can ``collapse'' into different possible outcome (post-measurement) states: $\psi_1, \psi_2, \psi_3,\dots$, but the experimenter cannot say in advance, not even in principle, which of them will be selected (see Figure~\ref{Figure1}).

However, a powerful rule exists in quantum mechanics, called de \emph{Born rule}, providing the probabilities for the different outcomes, linking in this way the theory with the experiments. And the so-called \emph{measurement problem} in quantum mechanics, in a nutshell,  is about finding the missing description of what goes on, ``behind the scenes,'' during a measurement process, also explaining the origin of the quantum probabilities and their values, as predicted by the Born rule. 

A possible solution to the measurement problem consists in taking the quantum fromalism very seriously, in the sense that: (1) the state of the physical system is assumed to describe its actual condition, and not just our knowledge of it; (2) a quantum measurement is assumed to be an objective physical process of change of state, and not a subjective process of acquisition of knowledge on the part of the experimenter; (3) the experimenter's consciousness is assumed to play no causal role (no psychophysical effects); (4) a measurement can only produce a single outcome at a time, and not all possible outcomes at the same time, each in a different ``parallel world.''

In other terms, we make here the effort of trying to identify a physical mechanism able to explain how the quantum probabilities can result from the interaction between the measuring apparatus and the measured system. For this, it is useful to come back to the \emph{absolute principle of the experimental method}. It was enounced in 1949 by \emph{Claude Bernard}, the father of scientific physiology, and it affirms that: ``If an experiment, when repeated many times, gives different results, then the associated experimental conditions must have been different each time.''

Because of their classical training, physicists were initially brought to assume that what could be different each time was the initial state of the physical system, and that by taking these state-fluctuations into account, the quantum probabilities could be explained. This, however, did not work as expected, because of the so-called \emph{no-go theorems}, an example of which are the famous \emph{Bell's inequalities}. Another possibility, proposed by the Belgian physicist \emph{Diederik Aerts}, in the eighties of the last century \cite{Aerts1986}, is to attribute the fluctuations not to the states of the measured system, but to its possible interactions with the measuring apparatus. Then, the no-go theorems no longer apply, and it becomes possible not only to conceptually explain the nature of a quantum measurement, but also to derive (in a non-circular way) the Born rule.

Aerts' proposal is known as the \emph{hidden-measurement interpretation} (HMI) of quantum mechanics, where the term ``hidden-measurement'' refers to the different measurement-interactions that would be available to be actualized at each run of a quantum measurement, explaining why for a same initial state different outcomes can be obtained. In recent times, the approach received a more complete treatement, in what was called the \emph{extended Bloch representation} (EBR) of quantum mechanics. The purpose of the present article  is to show how this representation works and why it is a possible solution to the quantum measurement problem. One of its advantages, as we will see, is that it also allows to visualize the unfolding of a quantum measurement, in a clear and unexpected way. 

Regarding the problem of visualizing a quantum process, the American physicist \emph{Jeremy Bernstein} wrote: ``Once he had his matrix mechanics, Heisenberg was able to reproduce all the results of the old quantum theory and more. It was the first example of a kind of Faustian bargain quantum theorists were to make with the spirit of visualization -- namely, one would be allowed to predict experimental results with very high accuracy provided that one did not ask for a visualization of the phenomena that went beyond the rules themselves.'' The German physicist \emph{Max Born}, the discoverer of the statistical rule that bears his name, also wrote: ``No language which lends itself to visualizability can describe quantum jumps.'' And the English physicist \emph{Paul Dirac} was of the opinion that: ``In the case of atomic phenomena no picture can be expected to exist in the usual sense of the word `picture'.'' This idea, that quantum processes would be beyond the reach of any explicit visual representation, is supported by many physicists even today. However, as we are going to show, it is the result of a misconception. Indeed, quantum measurements can be easily visualized, and therefore explained, using the EBR and its HMI.  

We start by describing the simplest possible situation: that of a measurement that can only produce two different outcomes. This is the case for instance when we measure the \emph{spin} of an electron (its intrinsic \emph{angular momentum}) along a given spatial direction, whose values can only be found to be either $+{\hbar\over 2}$ or $-{\hbar\over 2}$ (and for this reason it is called a spin one-half). For this, it is important to know that in 1946, the Swiss physicist \emph{Felix Bloch} was able to show that one can always represent the states of a two-dimensional quantum system (like a spin one-half system) as points at the surface of a three-dimensional sphere, of unit radius, nowadays called the \emph{Bloch sphere}.

\begin{figure}
\begin{center}
\includegraphics[width=7cm]{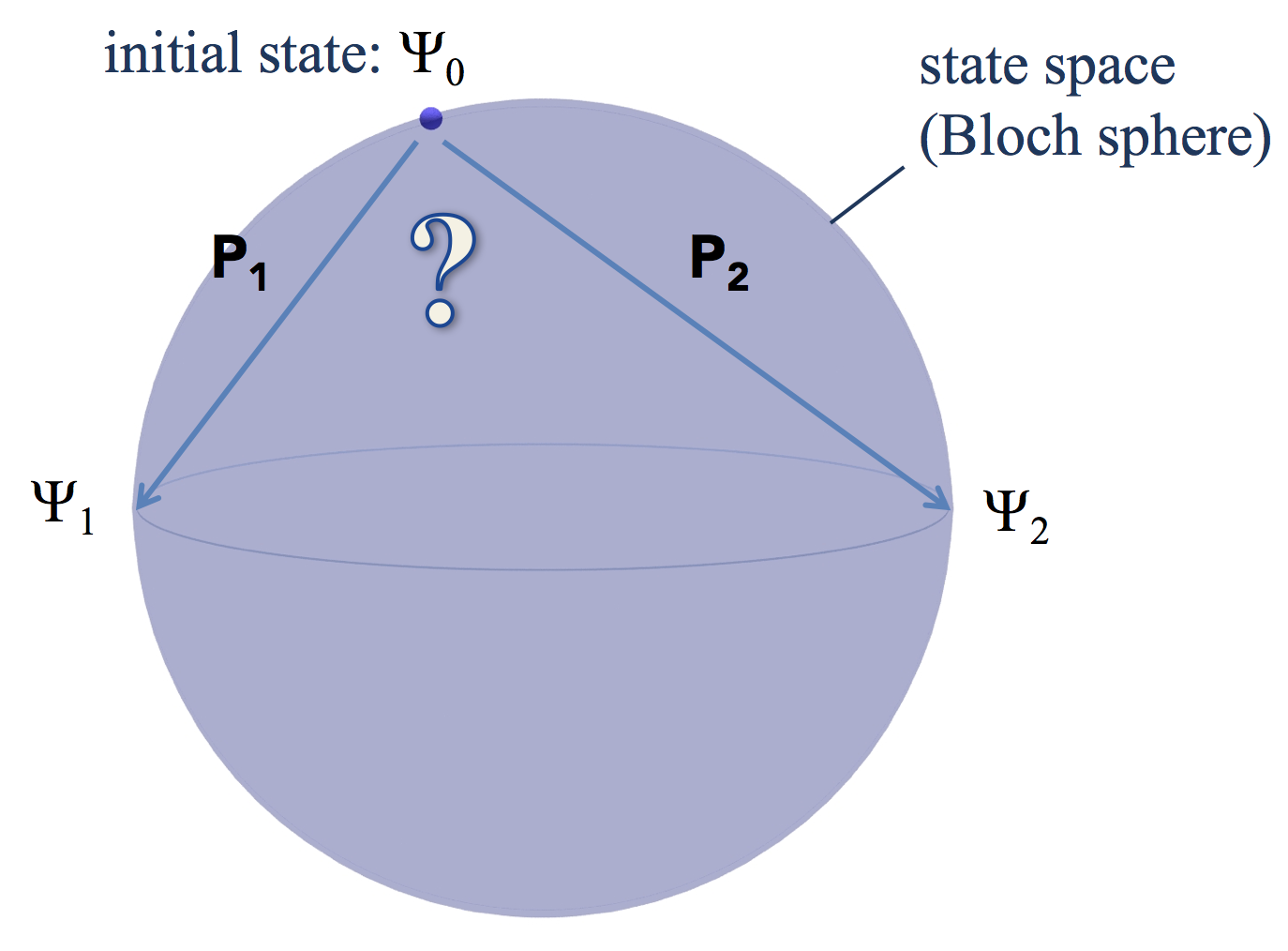}
\caption{The states of a two-dimensional quantum system can be represented as points at the surface of a three-dimensional unit sphere, called the Bloch sphere. A  measurement produces the transition from the initial state $\psi_0$ to either the outcome-state $\psi_1$, with probability $P_1$, or the outcome-state $\psi_2$, with probability $P_2$, where $\psi_1$ and $\psi_2$ are two antipodal points of the sphere.} 
\label{Figure2}
\end{center}
\end{figure}

This means that one can represent the initial state $\psi_0$ of the system, prior to the measurement, as an abstract point particle at the surface of the Bloch sphere, and same for the two outcome-states of the measurement, $\psi_1$ and $\psi_2$, which can be shown to necessarily be \emph{antipodal} points. So, in this Bloch sphere representation, the measurement problem is about finding a mechanism that can explain how an abstract point particle, representative of the initial state $\psi_0$, can non-deterministically move either to the state $\psi_1$, or to the state $\psi_2$, in accordance with the Born rule (see Figure~\ref{Figure2}). This means that if we repeat many times the process, the obtained experimental probabilities for the two transitions have to correspond to those theoretically predicted by the Born rule. 

To do so, we need to extend the Bloch sphere representation, to be able to represent not only the states, but also the measurement-interactions. For this, we observe that, by definition, two antipodal points of the sphere are so situated that a line drawn from one point to the other will pass exactly through the center of the sphere, and form a \emph{true diameter}. This true diameter describes the \emph{potentiality region} that characterizes the \emph{quantum observable} (like the spin-${1\over 2}$ observable of an electron) which has the two states $\psi_1$ and $\psi_2$ as its outcome-states (called \emph{eigenstates} in the quantum jargon). A simple way to describe and visualize the action of this potentiality region, is as an abstract uniform, sticky and breakable elastic band, stretched between the two antipodal points $\psi_1$ and $\psi_2$. But it is now time to see how a quantum measurement unfolds, in the EBR.

\begin{figure}
\begin{center}
\includegraphics[width=12cm]{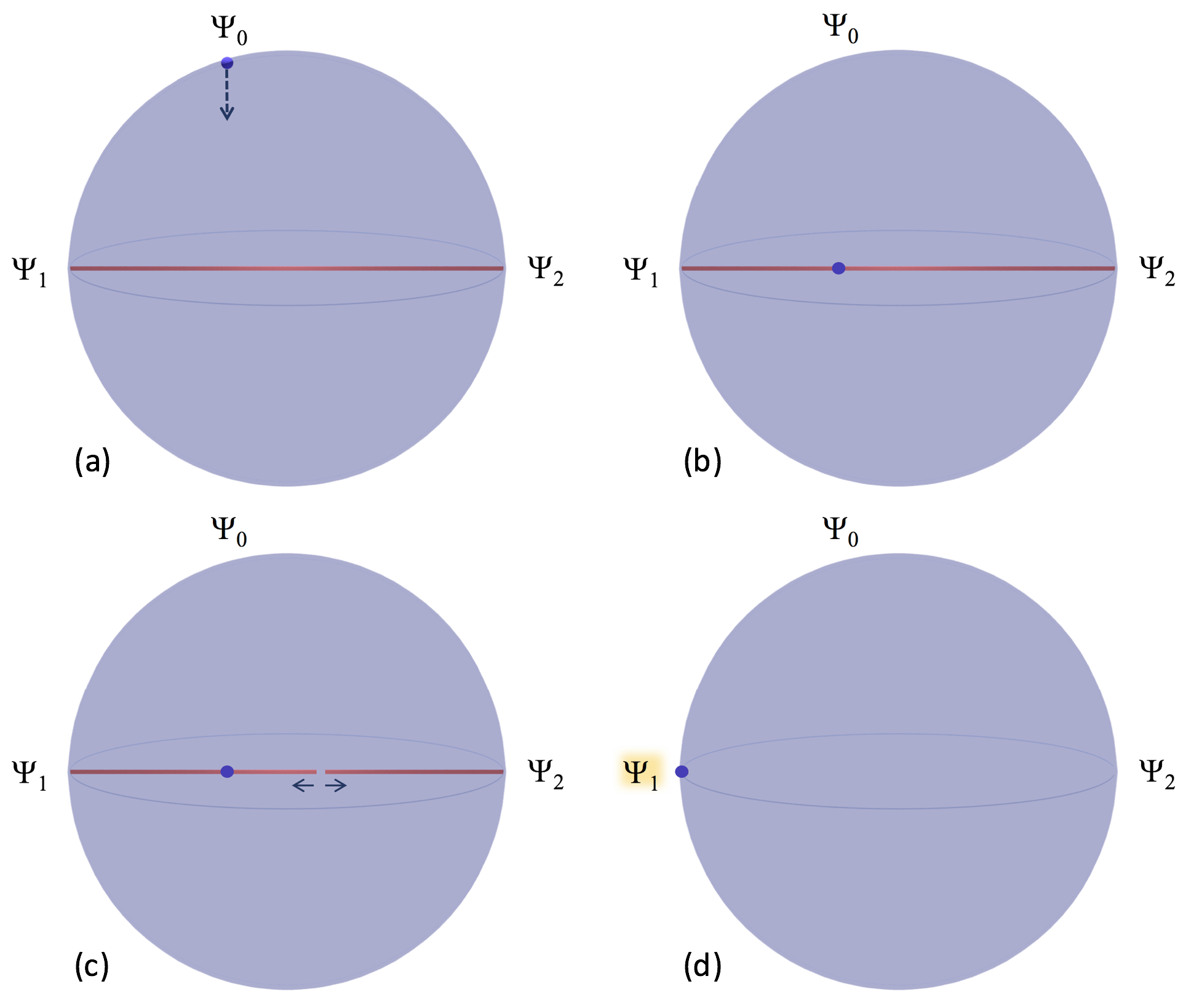}
\caption{A two-outcome quantum measurement: (a) the point particle representative of the initial state $\psi_0$ is located somewhere at the surface of the Bloch sphere, and then pluges into it along a trajectory that is orthogonal to the elastic band; (b) by doing so, it lands on the latter and firmly attaches to it; (c) when the elastic breaks, at some unpredictable point, by collapsing it brings the point particle to one of the two outcome-states, here $\psi_1$.} 
\label{Figure3}
\end{center}
\end{figure}

So, we have a Bloch unit sphere, which as we said is the sphere of states. We have an abstract point particle on it, representative of the initial state $\psi_0$ of the measured quantum system. And we have an abstract one-dimensional elastic structure, representative of the physical observable that we want to measure, with its two end points corresponding to the two outcome-states (if the observable is the spin of an electron, then the different possible orientations in the sphere of the elastic band correponds to the diffetent possible spatial directions along which the spin can be measured, for instance by means of a Stern-Gerlach apparatus). The measurement then corresponds to the following two-stage process: (1) during the first stage, which is deterministic, the abstract point particle plunges into the sphere along a path that is orthogonal to the elastic band, and firmly attaches to it; (2) during the second stage of the measurement, which is non-deterministic, the abstract elastic band breaks at some unpredictable point, and when this happens the point particle, being attached to one of the two broken fragments, is pulled towards one of the two end points, producing in this way the final outcome of the measurement (see Figure~\ref{Figure3}).

Clearly, depending on where the abstract elastic structure breaks, one rather than the other outcome will be obtained. More precisely, each breakable point of the elastic band corresponds to a different possible interaction between the latter and the point particle attached to it. The breakable elastic band is therefore representative of the collection of all the potential \emph{measurement-interactions} that can be actualized during the quantum measurement process. Let us also mention that when the point particle ``falls'' onto the elastic, the physical systems experiences a so-called \emph{decoherence} process, which can be understood as its entanglement with the environment. Decoherence, however, is not sufficient to produce an outcome, which can only be obtained when the elastic breaks, causing the point particle to be drawn to one of the two possible end-states. 

\begin{figure}
\begin{center}
\includegraphics[width=5.5cm]{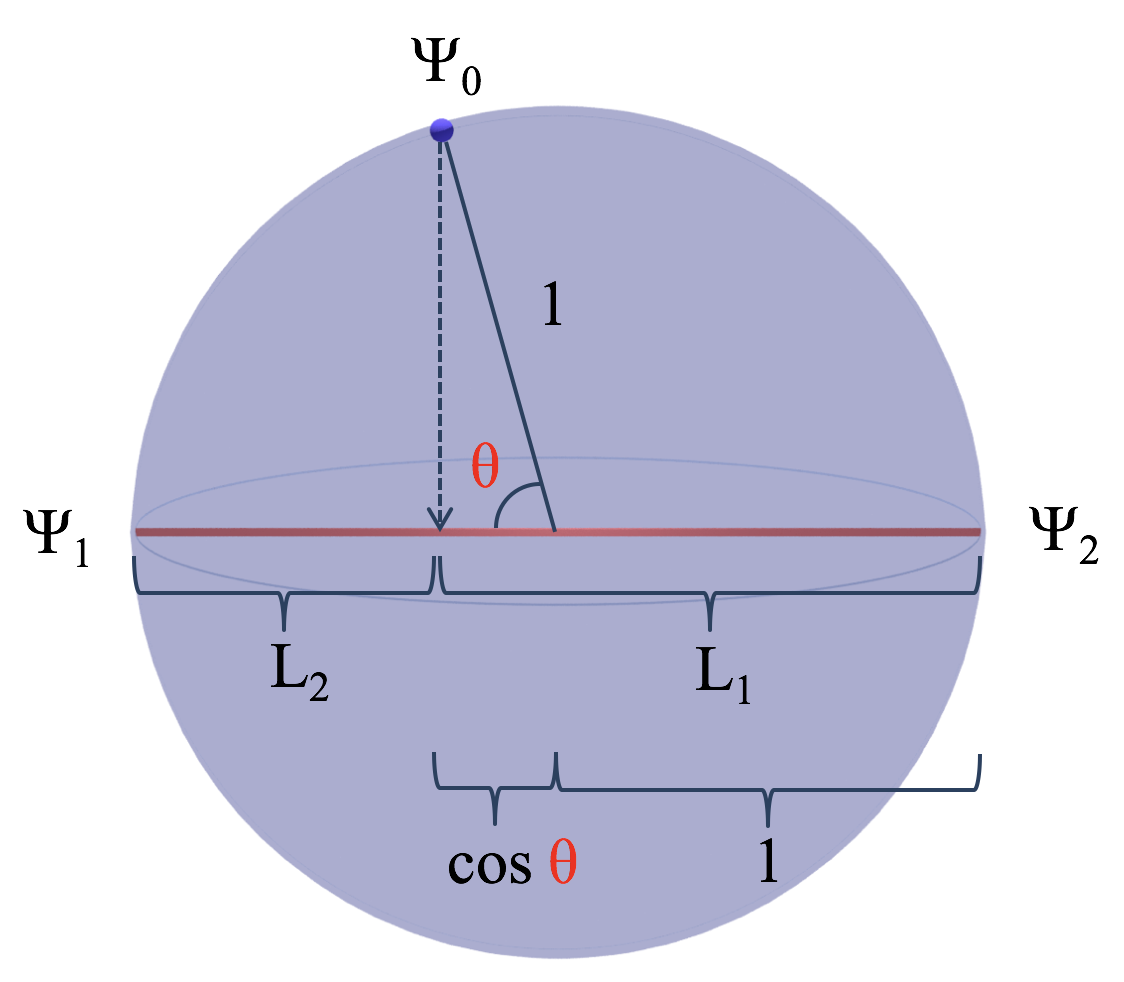}
\caption{When the point particle lands onto the elastic band, two line segments $L_{1\atop 2}=1\pm\cos\theta$ are obtained.} 
\label{Figure4}
\end{center}
\end{figure}

Let us show that the process we have just described allows to obtain the exact quantum probabilities, that is, to derive the Born rule. For this, we observe that the point particle representative of the initial state $\psi_0$ makes a certain angle with respect to the elastic band. Let us call it $\theta$. We also observe that when the particle ``falls'' onto the elastic, it defines two line segments: $L_1$ and $L_2$ (see Figure~\ref{Figure4}). It is clear that if the elastic breaks in $L_1$, the final outcome will be $\psi_1$, and if breaks in $L_2$, the final outcome will be $\psi_2$. Also, being by hypothesis the elastic uniform, and being its total length $2$ (twice the unit radius of the sphere), the probability $P_1$ for the transition to $\psi_1$ is given by the length of the segment $L_1$ divided by $2$. And considering that the length of$L_1$ is $1+\cos\theta$, we have: $P_1={L_1\over 2}={1+\cos\theta\over 2}=\cos^2{\theta\over 2}$, and a similar reasoning gives $P_2=1-P_1=\sin^2{\theta\over 2}$, which are exactly the expressions predicted by the Born rule, for a two-outcome measurement situation.

At this point, the typical objection is that if we are able to derive the Born rule in this way, it is just because we have limited our discussion to two-dimensional (two-outcome) quantum systems. Indeed, the celebrated \emph{Gleason's theorem}, which was instrumental in ruling out the existence of hidden-variable explanations of quantum theory, is known to be valid only for situations with more than two outcomes. Note that the term ``hidden-variables'' refers here to those elements of reality about which we lack knowledge and which would explain why we can only predict the outcomes of a measurement in probabilistic terms. In other words, one may suspect that the possibility of a HMI would be just a two-dimensional anomaly, impossible to generalize to experimental situations with more than two-outcomes. 

This objection is however unfounded, and this for at least two reasons. The first one, which we mentioned already, is that in the EBR the hidden-variables are not referring to the states of the measured system, but to the measurement-interactions. This is why their introduction does not restore determinism, which is what physicists have historically tried to achieve when exploring hidden-variables models. In other words, the no-go theorems, like Gleason's theorem and Bell's inequalities, only forbid the replacement of quantum mechanics by a more fundamental fully deterministic theory in which the probabilities of having or not having a certain property can only take the values $0$ or $1$.  The second reason is that the EBR can be naturally generalized to more general measurement situations, with arbitrary number of outcomes (and in fact, even to so-called degenerate measurement situations, where different outcome-states are associated with the same value of the quantum observable). 

\begin{figure}
\begin{center}
\includegraphics[width=15cm]{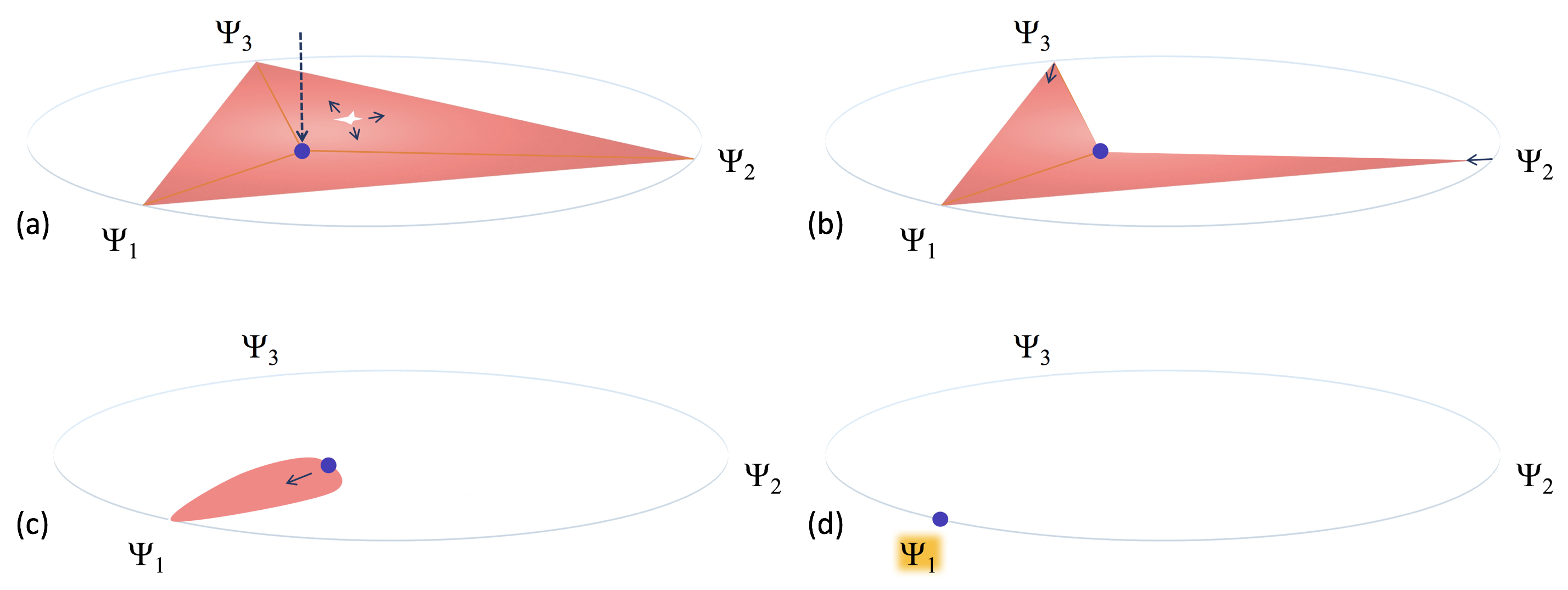}
\caption{A three-outcome quantum measurement: (a) the point particle representative of the initial state $\psi_0$, initially located at the surface of the generalized eight-dimensional Bloch sphere,  plunges into it along a trajectory orthogonal to the elastic membrane, firmly attaching to it, which then starts disintegrating at some unpredictable point; (c) the disintegration propagates in the subregion where it started, causing the detachment of two of the three anchor points of the membrane; (c) the latter then contracts towards the only remaining anchor point; (d) by doing so, it brings the point particle to that same point, here $\psi_1$.} 
\label{Figure5}
\end{center}
\end{figure}

So, our next step is to describe the unfolding of a quantum measurement having three different possible outcomes (for example, the measurement of a spin-$1$ observable along a given direction, whose values can only be found to be $\hbar$, $0$ and $-\hbar$). It can be shown that the initial state $\psi_0$ of the system, prior to the measurement, is now described by an abstract point particle at the surface of a eight-dimensional generalized Bloch sphere, which of course we can no longer explicitly draw. To put it plainly, different from the two-outcome case, only a convex portion of such generalized Bloch sphere is filled with \emph{bona fide} states, but this is a technical detail we don’t need to bother here. The three outcome-states, let us call them $\psi_1$,  $\psi_2$ and $\psi_3$, can of course also be represented as points located at the surface of this 8-dimensioanal hyper-sphere, and it is possible to demonstrate that they necessarily lay on a same two-dimensional plane, and that they correspond to the vertices of an equilateral triangle inscribed in the sphere (see Figure~\ref{Figure5}). 

Again, we want to describe how the abstract point particle representative of the initial state $\psi_0$ can non-deterministically transition to one of the three final states, in accordance with the Born rule, which means that if we repeat many times the process, then the experimental probabilities $P_1$, $P_2$ and $P_3$, for the three possible transitions, have to correspond  to those theoretically predicted by the Born rule.  For this, we now describe the potentiality region associated with the measured quantum observable, and delineated by the three outcome-states $\psi_1$,  $\psi_2$ and $\psi_3$, as a uniform,  sticky and disintegrabile elastic membrane, stretched between these three points representative of the outcomes. But let us see how the quantum measurement unfolds in this case (see Figure~\ref{Figure5}). 

So, we have a generalized eight-dimensional unit Bloch sphere, and we have an abstract point particle on it, representative of the initial state $\psi_0$ in which the quantum system was prepared. And we have a two-dimensional abstract triangular membrane, representative of the observable that we want to measure, with its three vertices corresponding to the three possible outcomes of the measurement. Then, the measurement corresponds to the following two-stage process: (1) similarly to the two-outcome situation, during the first deterministic stage the point particle plunges into the hyper-sphere, along a path that is orthogonal to the triangular membrane, and firmly attaches to it. When this happens, it defines three different triangular subregions on the membrane, distinguished by line segments connecting the on-membrane particle's position with the three vertex points. We have to think of these line segments as ``tension lines,'' making the membrane less easy to disintegrate along them; (2) the second stage of the measurement, purely non-deterministic, goes as follows: once the particle is attached to the membrane, the latter start disintegrating at some unpredictable point, belonging to one of the three subregions. The disintegration then propagates inside that specific subregion, but not in the other two, because of the tension lines. This causes the two anchor points of the disintegrating subregion to tear away, so producing the detachment and subsequent collapse of the elastic membrane, which by contracting towards the only remaining anchor point draws to that same position the point particle, which reaches its final state, corresponding to the outcome of the measurement (see Figure~\ref{Figure5}). 

\begin{figure}
\begin{center}
\includegraphics[width=5cm]{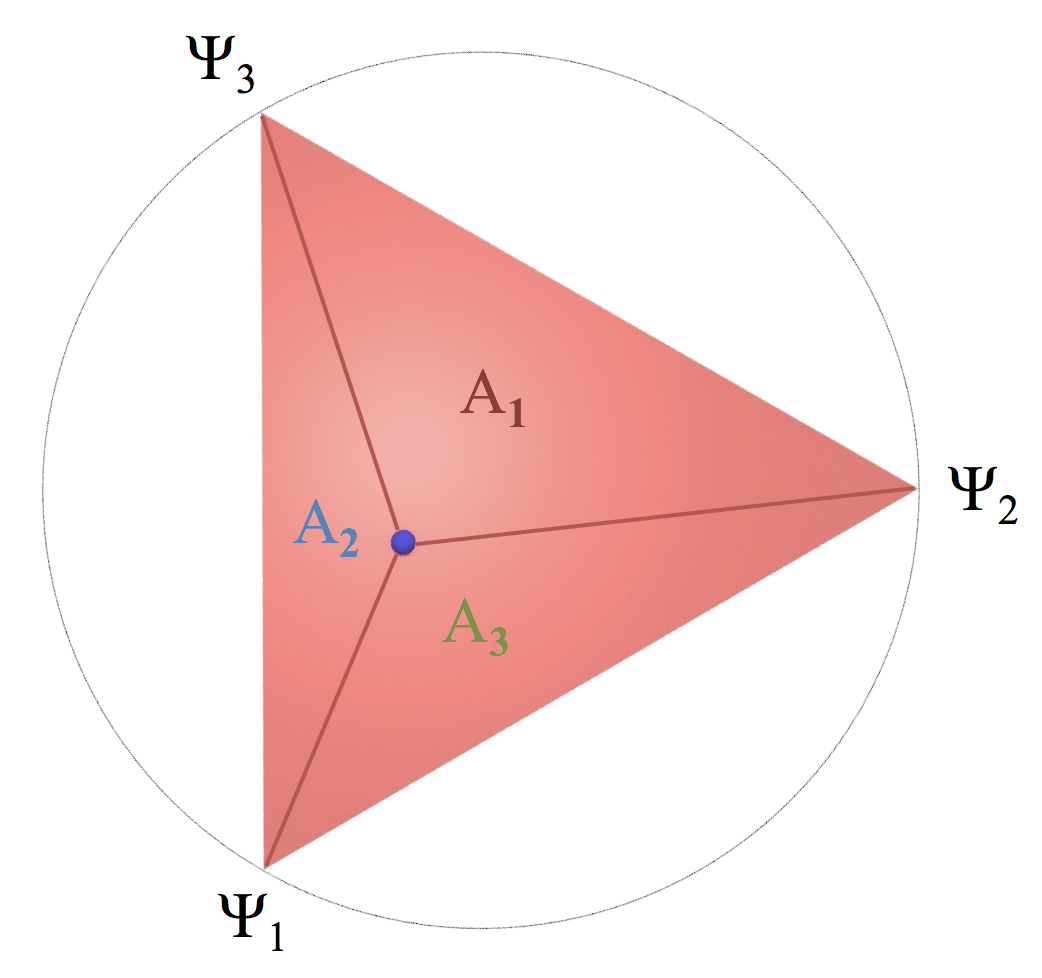}
\caption{The three regions obtained when the point particle lands on the elastic triangular membrane, whose areas depend on the orientation of the initial state with respect of the outcome-states.} 
\label{Figure6}
\end{center}
\end{figure}

As for the two-outcome situation, the next step is to show that that the process we just illustrated allows to derive the exact quantum probabilities (\emph{i.e.}, the Born rule. As we observed, the point particle, when attached to the membrane, defines three distinct sub-regions, $A_1$, $A_2$ and $A_3$ (see Figure~\ref{Figure6}). If the membrane starts disintegrating in $A_1$, the final outcome will be $\psi_1$, if it starts disintegrating in $A_2$, the final outcome will be $\psi_2$, and if it starts disintegrating in $A_3$, the final outcome will be $\psi_3$. Being the membrane, by hypothesis, uniform, the probability $P_1$ for the transition to $\psi_1$ is given by the area of the sub-region $A_1$ divided by the total area of the membrane (which is ${3\sqrt{3}\over 4}$ for a rectangular trinagle inscribed in a unit sphere), and similarly for the other two transitions. 

If the calculation of these relative areas is done properly, it is possible to show that they reproduce exactly the probabilities predicted by the Born rule. Of course, we will not perform these calculations here, as this would lead us to a too technical discussion that is beyond the scope of the present article, and we refer the readers with knowledge of the quantum mathematics to \cite{AertsSassolideBianchi2014}, where the proof is worked out for the general situation of a quantum mesurement with an arbitrary number $N$ of outcomes. The generalized Bloch sphere is then ($N^2-1$)-dimensional and the potentiality region is a $(N-1)$-dimensional \emph{simplex} inscribed into it (a simplex is a figure generalizing the notion of an equilateral triangle to higher dimensions; for $N=4$, it corresponds for example to a \emph{tetrahedron}). 

Much more should be said about the EBR and its HMI, but it will be sufficient to conclude with a few remarks. What the EBR reveals is that, incorporated in the quantum formalism, there is a full description of the measurements processesses, as unstable equilibrium situations where different measurement-interactions mutually compete in the actualization of a specific outcome. Quantum probabilities can therefore be understood as \emph{epistemic} statements, expressing a condition of lack of knowledge about which specific measurement-interaction is selected at each run of a measurement. At the current state of our knowledge, we don't know if a solution to the measurement problem in terms of hidden measurement-interactions is the correct one, \emph{i.e.}, if these non-ordinary interactions are truly present in our physical reality. But certainly this is a challenging theoretical solution, awaiting for a possible future experimental confirmation. 

It is also worth observing that the EBR allows for the description of more general measurements than just the pure quantum ones, \emph{i.e.}, measurements characterized by non-uniform breaking/disintegrating elastic structures. Interestingly, it can be shown that when a huge average over all possible kinds of non-classical (non-uniform) measurements is considered -- called a \emph{universal average}~\cite{AertsSassolideBianchi2017} -- one finds back again the Born rule \cite{AertsSassolideBianchi2014}. This means that quantum probabilities admit an interpretation not only as resulting from a situation of lack of knowledge about which measurement-interaction is each time actualized, but also about how such actualization process is each time implemented.
\\

\noindent {\it Further readings}. A very accessible analysis of the EBR and HMI can be found in \cite{AertsSassolideBianchi2017}, as well as its relation to other fundamental  probems, like Bertrand's paradox \cite{AertsSassolideBianchi2014b} and the ``unreasonable'' success of quantum probabilities beyond the domain of physics \cite{AertsSassolideBianchi2015a}. An dialogue confronting the many-measurements with the many-worlds interpretations, showing that the former allows to deduce an even more fantastic scenario than that of the multiverse, can be read in \cite{AertsSassolideBianchi2015b}. In \cite{AertsSassolideBianchi2017b}, the EBR and HMI are also compared with the \emph{transactional interpretation}, particularly concerning the possibility of understanding a quantum measurement as weighted symmetry breaking process. Let us mention that the EBR allows to clarify and demistify other quantum phenomena, like entanglement, showing that it remains compatible with the principle according to which a composite entity exists only if its components also exist, and therefore are in well-defined states \cite{AertsSassoli2016b,AertsSassoliSozzo2017}. Finally, we point out that some nice computer animations of the unfolding of a measurement in the EBR, for the $N=2,3,4$ situations, can be found in \cite{SassoliSassoli2015}.

\end{document}